\documentclass[aps,prl,reprint,groupedaddress,floatfix]{revtex4-1}
\usepackage{graphicx}
\usepackage{xcolor}
\usepackage{soul}
\begin{document}

\title{Search for dark photon, axion-like particles, dark scalar, or light dark matter in Compton-like processes}
\author{Sankha S. Chakrabarty and Igal Jaegl\'e \\Contact~e-mail: \href{mailto:igjaegle@gmail.com}{igjaegle@gmail.com} }
\affiliation{Department of Physics, University of Florida, Gainesville 32611, Florida, USA}

\date{\today}

\begin{abstract}

We propose a novel way to search for the dark photon ($A'$), the axion-like pseudo-scalar ($a$), the dark scalar ($\phi$), and the light dark matter ($\chi$) in the Compton-like processes, $\gamma + e^- \rightarrow A'/a/\phi + e^-$ with $A'/a/\phi$ decaying into leptons, photons, or $\chi$'s (when permitted) for the mass ranges of
  $m_{A'/a/\phi} \leq$ 100 MeV/$c^2$ and $m_{\chi} \leq $ 50 MeV/$c^2$. We examine how the past, current, and future fixed target experiments can use this under-explored production mechanism of dark particles. We show that the existing and planned tagged photon beam fixed target experiments (GlueX, LEPS2, LEPS, and FOREST) are competitive with the electron beam fixed target experiments, particularly for searching an invisible particle. We also show that the photon flux produced in the beam dump experiments increases the dark particle flux significantly and therefore the sensitivities of these experiments. By only considering the Compton-like processes, we determine the new limits and expected sensitivities for several beam dump and active beam dump experiments (E774, E141, KEK, Orsay, E137, NA64, BDX and LDMX) on the kinetic mixing parameter ($\epsilon$) between the dark photon and the Standard Model photon, the axion-like pseudo-scalar coupling to electrons ($g_{ae}$), the dark scalar coupling to electrons ($y_e$) and the dimensionless interaction strength ($y$) to light dark matter.    
  
\end{abstract}


\maketitle

\section{Introduction}
\label{S:1}

\begin{figure}[t!]
\begin{center}
\includegraphics[scale=0.5]{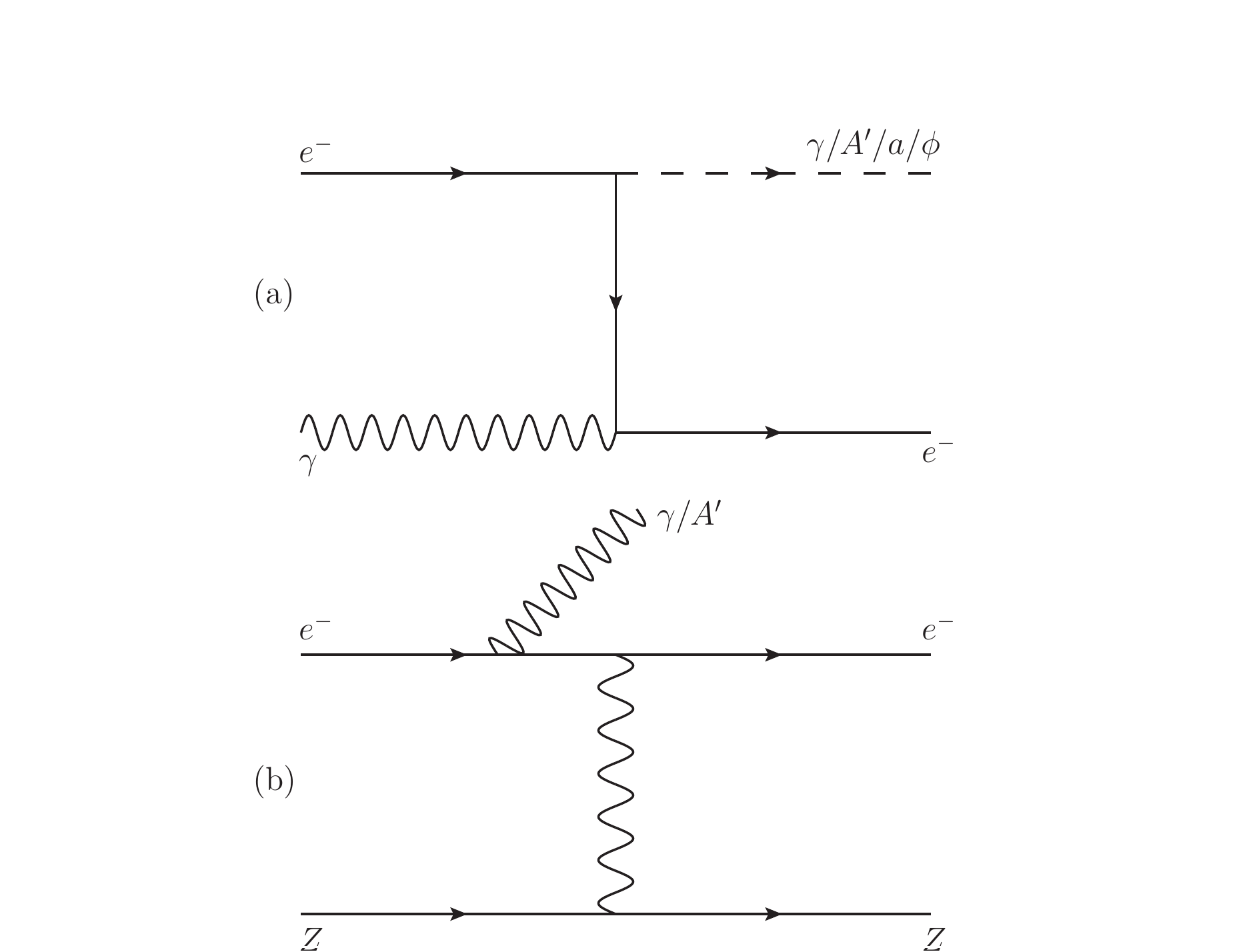}
\caption{Possible $A'$ production modes in fixed target photon, lepton, or hadron beam experiments: (a) $A'$($/\gamma/a/\phi$)-Compton and (b) $A'$($/\gamma$)-bremsstrahlung.}
\label{fig:FeynmanDiagrams}
\end{center}
\end{figure}

The Universe missing mass ``problem" was observed for the first time in 1933~\cite{Zwicky:1933,Zwicky:1937}, and has been reported since then at various scales and ages of the Universe~\cite{Rubin:1980ApJ,Clowe:2006ApJ,2016PhRvL.117t1101M,Harvey:2015hha,Ade:2015xua}. These cosmological anomalies can be explained by non-luminous matter (the so-called Dark Matter DM), or a modification of the theories of gravity~\cite{Milgrom:1983ApJ}, or a combination of both DM and modified gravity~\cite{Famaey2012}. Recently, the EDGES Collaboration~\cite{Bowman:2018} reported a lower than expected baryon temperature during the time period between 190~Myr and 240~Myr after the Big Bang. This anomalous temperature can be caused by the presence of light dark matter~\cite{Barkana:2018}, light milli-charged particle~\cite{Barkana:2018qrx,Munoz:2018pzp,Kovetz:2018zan,Berlin:2018sjs}, QCD axion, or axion-like particles~\cite{Sikivie:2018tml,Kovetz:2018zan,Berlin:2018sjs}. The search strategies for these hypothetical particles in space- and Earth-bound experiments depend on the assumptions about how these particles can be produced and how they interact with the Standard Model particles. The Axion Models~\cite{Redondo:2013wwa} are proposing a broad spectrum of processes (hadronic and non-hadronic) to use for these searches and are referring to the non-hadronic processes as A-B-C-P-reactions (Atomic recombination and de-excitation, Bremsstrahlung, Compton, and Primakoff) in contrast to the Dark Sector Models (DSM)~\cite{Bjorken:2009mm} which are almost exclusively suggesting to use the bremsstrahlung-like processes or B-reactions (Figure~\ref{fig:FeynmanDiagrams}-b). No signal or hint has been found so far by the lepton/hadron beam experiments. The ATOMKI Collaboration~\cite{Feng:2016,Feng:2016ysn} reported an anomalous excess of $e^+e^-$ pair at 16.7~MeV/${\it c}^2$ in the excited $^8$Be$^*$ nucleus decay which is an Atomic de-excitation or A-reaction, an experimentally understudied process. In this paper, we are discussing another experimentally understudied process,  the Compton-like process or C-reaction (Figure~\ref{fig:FeynmanDiagrams}-a), $\gamma~e^- \rightarrow (A'/a/\phi) ~e^-$, which is relevant in four scenarios: (1) the neutrino experiments ~\cite{Park:2017prx}, (2) a future lepton-photon collider~\cite{Alesini:2013szx,Gakh:2018} or electron-laser experiments~\cite{Brodsky:1986}, (3) the tagged photon beam fixed target experiments and (4) the beam dump and active beam dump experiments. Here, we are particularly focusing on the last two scenarios.

\section{Theory}
\label{S:2}

For the Compton-like process to produce dark photon ($A'$): $\gamma~e^- \rightarrow A'~e^-$, the interaction of an electron ($\psi_e$) with dark photon ($A'$) is described by the following term in the Hamiltonian:
\begin{equation}
\Delta H = \int d^3 x \; \epsilon e \; \bar{\psi}_e \gamma^{\mu} \psi_e A'_{\mu} \label{Hintvector}
\end{equation}
where $e$ is the electronic charge and $\epsilon$ is the kinetic mixing parameter between the dark photon and the Standard Model photon~\cite{Fayet:1980,Holdom:1985ag,Fayet:1986,Boehm:2003hm,Fayet:2004bw,Pospelov:2008zw}. The exact (free) differential cross section for this process, $\frac{d\sigma}{d\Omega}$, is given in the Appendix. 
The total cross-section for this process is approximately given by
\begin{equation}
\sigma \approx \frac{4\pi \alpha^2 \epsilon^2}{s} \text{ln}\Big(\frac{1 - x_M^2}{x_m}\Big) \label{sigmatotA}
\end{equation}
where $\alpha = \frac{e^2}{4\pi}$ is the fine structure constant, $\sqrt{s} = E_{\text{com}}$ is the energy in center of momentum frame, $m_e =x_m\sqrt{s}$ is the electron mass and $M = x_M\sqrt{s}$ is the dark photon mass. The Standard Model Compton cross-section can be obtained by putting the dark photon mass to be zero and the kinetic mixing to be unity ($x_M = 0$ and $\epsilon = 1$ in Eq.~\ref{sigmatotA}), as expected. For a photon beam with energy $E_{\gamma}$ hitting an electron at rest in the lab frame, $\sqrt{s} = \sqrt{2m_e E_{\gamma} + m_e^2}$. The threshold energy $E_T$ of the photon beam to produce a dark photon of mass $M$ is given by: 

\begin{eqnarray}\label{E_threshold}
E_T &&= M + \frac{M^2}{2m_e}.
\end{eqnarray}

Similar Compton-like processes can be used to produce axion-like pseudo-scalar particles ($a$ or ALPs), $\gamma e^- \rightarrow a e^-$, or dark scalar mediators ($\phi$), $\gamma e^- \rightarrow \phi e^-$. As, there is no virtual photon involved in the Compton-like processes, milli-charged particles cannot be produced. The relevant interaction terms are given by

\begin{eqnarray}
\Delta H &&= \int d^3 x \; g_{ae} \; \bar{\psi}_e \gamma_{5} \psi_e a \\
\Delta H &&= \int d^3 x \; y_{e} \; \bar{\psi}_e \psi_e \phi . \label{Hintscalars}
\end{eqnarray}
where $g_{ae}$ and $y_e$ are the ALP coupling to electron and the dark scalar coupling to electron, respectively. The total cross sections for the productions of ALPs and dark scalar mediators are given by~\cite{Brodsky:1986, Knapen:2017xzo}
\begin{eqnarray}
\sigma \; &&\approx \frac{\alpha g_{ae}^2}{4s} \left(2 \; \text{ln} \frac{1}{x_m} - \frac{3}{2}\right) \\
\sigma \; &&\approx \frac{\alpha y_{e}^2}{s} \left(\frac{5}{2} + \text{ln} \frac{1}{x_{m}^2 + x_{M}^2}\right) . \label{sigmatot}
\end{eqnarray}
The $A'$ can decay into leptons, hadrons, or light DM particles, $\chi$~\cite{Batell:2014mga}. The $a$ can decay into photons, electrons, and light DM particles. The $\phi$ can decay into photon, leptons, and light DM particles. When $A'(a/\phi) \rightarrow \chi\bar{\chi}$, the $A'(/a/\phi)$ coupling to $\chi$, $g_D = \sqrt{4\pi \alpha_D}$, is involved.

\begin{figure}[t!]
\begin{center}
\includegraphics[scale=0.5]{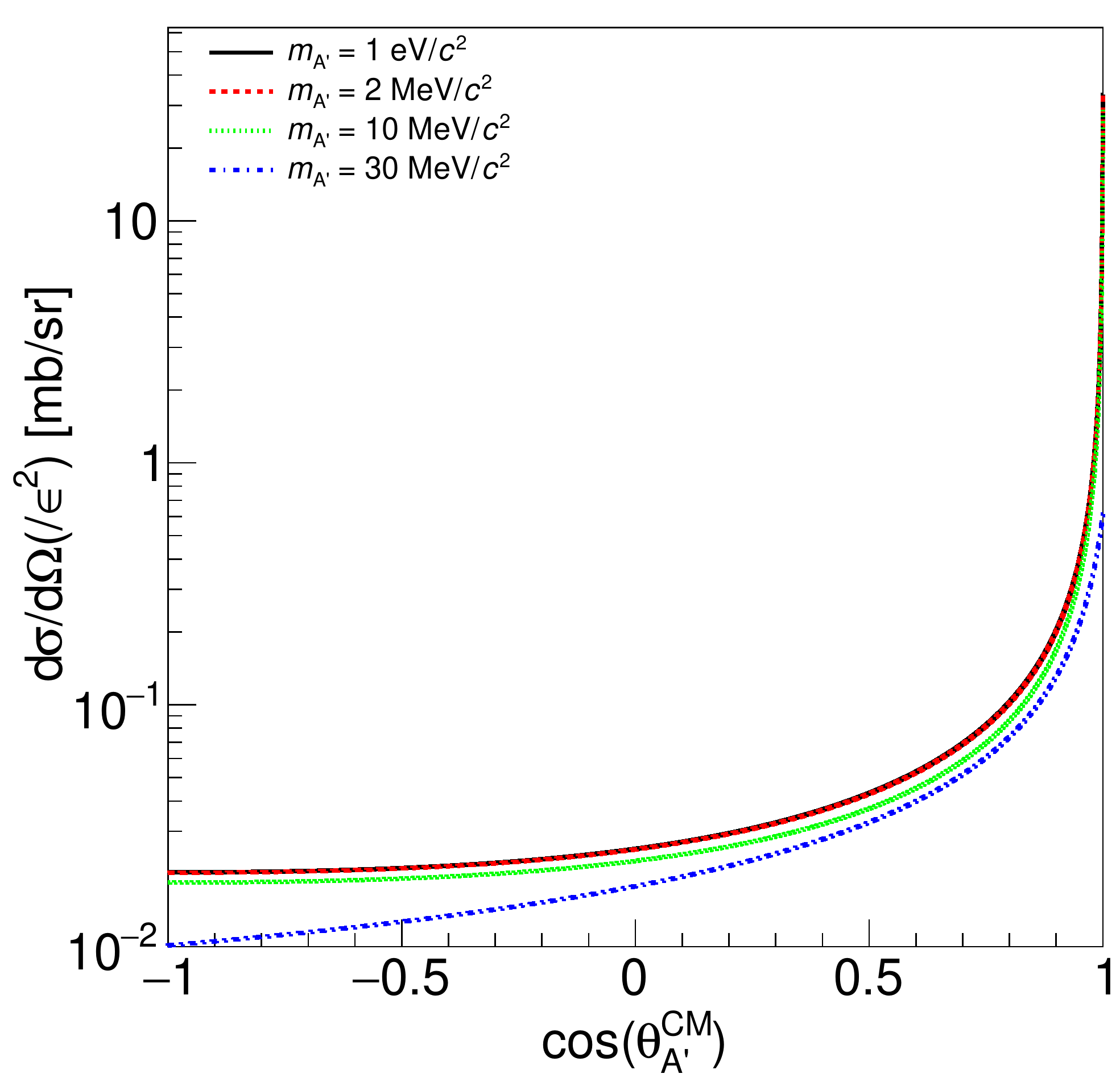}
\caption{Differential cross-section of the Compton-like process ($e^- \gamma \rightarrow e^- A'$) producing dark photon $A'$ of various masses $m_{A'}$ versus the dark photon polar angle in the COM frame. We have considered incident photon beam of 1~GeV hitting a free electron at rest.}
\label{fig:dxs}
\end{center}
\end{figure}

The numerical values of the total cross sections for a 10~GeV photon beam impinging a liquid Hydrogen target and producing a $10$~MeV/${\it c}^2$ ``dark" particle in the Compton-like processes are: 
\begin{eqnarray}
\sigma (e\gamma \rightarrow eA') &&\approx 1.4 \; \text{pb} \Big( \frac{\varepsilon}{10^{-4}} \Big)^2 \Big( \frac{0.1 \text{GeV}}{\sqrt{s}} \Big)^2 \label{NumA'} \\
\sigma (e\gamma \rightarrow ea) &&\approx 6.5 \; \text{pb} \Big( \frac{g_{ae}}{10^{-4}} \Big)^2 \Big( \frac{0.1 \text{GeV}}{\sqrt{s}} \Big)^2 \label{Numa} \\
\sigma (e\gamma \rightarrow e\phi) &&\approx 20.2 \; \text{pb} \Big( \frac{y_e}{10^{-4}} \Big)^2 \Big( \frac{0.1 \text{GeV}}{\sqrt{s}} \Big)^2 . \label{Numphi} 
\end{eqnarray}

Figure~\ref{fig:dxs} shows the differential cross-section for various dark photon masses versus the dark photon polar angle in the center of momentum (COM) frame. The differential cross-section for dark photon (as well as axion-like or dark scalar particle) production is strongly peaking in the same direction as the incident photon.

\begin{figure}
    \begin{center}
   
    \includegraphics[scale=0.45]{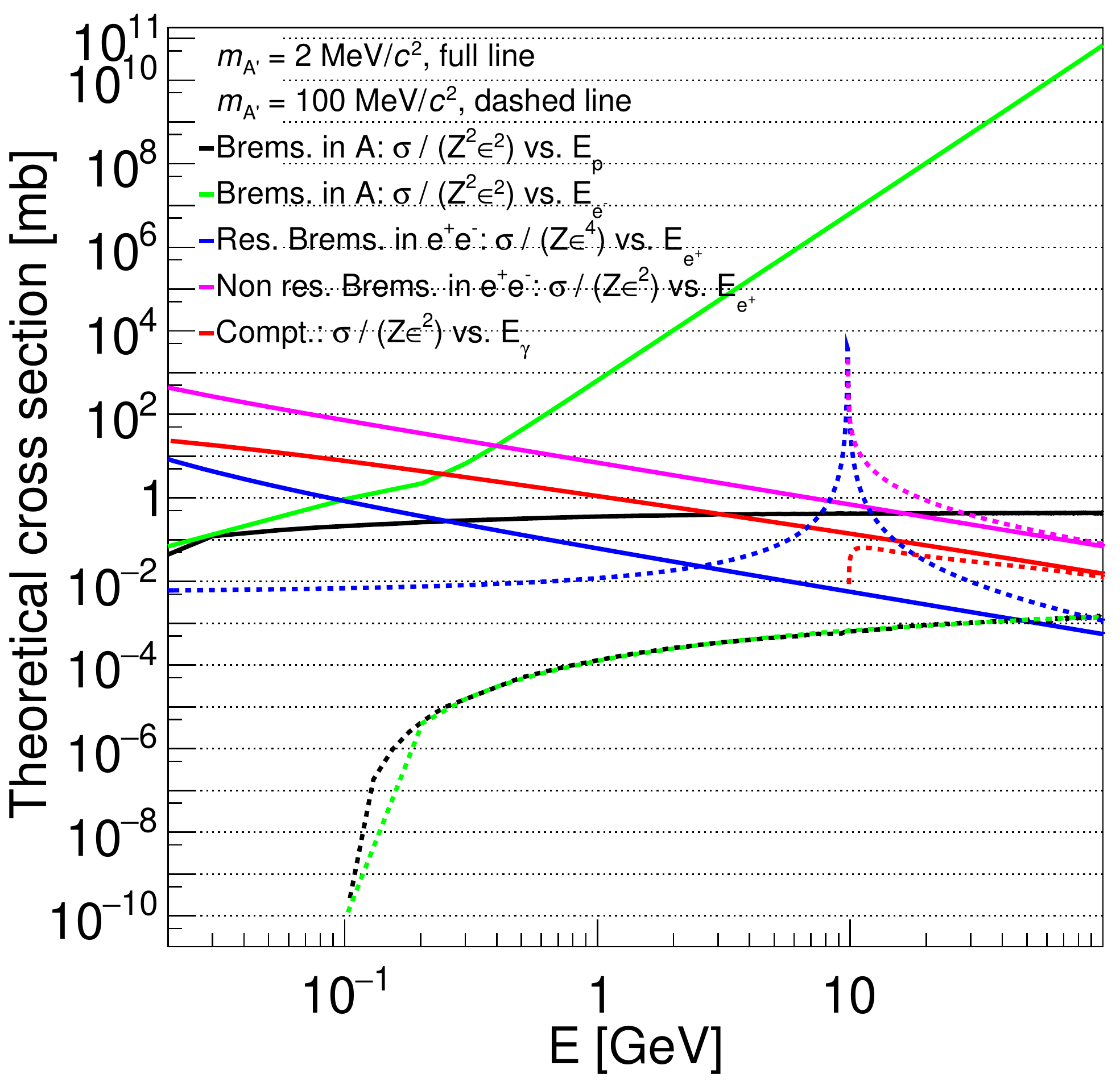}
    \end{center}
              \caption{Theoretical cross-sections (normalized to their dependencies to the kinetic mixing and the atomic number, when applicable) as functions of the  beam energy for two different dark photon masses and different processes. Solid and dashed lines for $m_{A'}$~=~2~MeV/${\it c^2}$ and 100~MeV/${\it c^2}$, respectively. Black and green curves for bremsstrahlung in nucleus scattering versus the proton- and electron-beam energy, respectively. Blue and purple curves for resonant and non-resonant bremsstrahlung in lepton-pair annihilation versus the positron-beam energy, respectively. Red curve for the Compton process versus the photon-beam energy.} 
               \label{fig:txs}
\end{figure}

Figure~\ref{fig:txs} shows the theoretical cross section versus energy of the incident beam of electron, positron, proton and photon. Cross-sections for  different processes are normalized to their atomic number $Z$ and kinetic mixing $\epsilon$ dependencies. The cross sections of Compton-like processes, and bremsstrahlung in lepton annihilation (both resonant and non-resonant) scale as $Z$ where $Z$ is the atomic number. For bremsstrahlung in nucleus scattering, the cross-sections scale as $Z^2$.  Therefore, for a 1~GeV incident beam energy and a 2~MeV/${\it c}^2$ dark photon mass, the theoretical cross-section of Compton-like process is $\sim$ 650($\times Z$) times smaller than the bremsstrahlung in nucleus scattering using an electron beam, 7 times smaller than the non-resonant bremsstrahlung in $e^+ e^-$ annihilation using a positron beam, but $\sim$18($/\epsilon^2$) times larger than the resonant bremsstrahlung using a positron beam and  $\sim$ 3($/Z$) times larger than the bremsstrahlung in nucleus scattering using a proton beam. It should be noted that the Compton-like cross-sections are independent of the dark particle mass if the incident photon beam energy is much larger than the production energy threshold.

Electrons are not free in fixed target experiments, but exist in the form of atomic electrons. Therefore atomic binding energy, screening and radiative corrections~\cite{Mork:1967,Maximon:1981,Dugger:2017} must be considered to evaluate a realistic cross section. Atomic binding energy is negligible as we are only considering photon energy above 20~MeV which is well above the K-shell binding energy. The corrected cross section can be written as:
\begin{equation}
    \sigma = \sigma_{\mathrm{free}}\; Z \left(1 + R \frac{\sigma(\gamma A \rightarrow Ae^+e^-)}{\sigma(\gamma e^-_A \rightarrow e^-_Ae^+e^-)}\right) \left(1 - F^2(q)\right) 
    \label{eqn:11}
\end{equation}
where $Z$ is the atomic number, $R = 0.0093$ is the $Z$-independent radiative correction, $F(q)$ is the Hydrogen form factor, and  $\sigma(\gamma A \rightarrow Ae^+e^-)$, $\sigma(\gamma e^-_A \rightarrow e^-_Ae^+e^-)$ are the cross-sections of the SM pair photoproduction off nucleon, the SM triplet photoproduction off electron, respectively. $F(q)$ can be expressed as: $F(q) = (1 - \frac{a^2q^2}{4})^{-2}$ where $a$ and $q$ are the Bohr radius and the momentum transfer to the recoil electron, respectively. 

\section{Experiments}

We review how the Compton-like processes can be used in three types of experiments: (1) tagged photon-beam fixed-target experiments ($E_\gamma E$), (2) beam-dump experiments (BDE), and (3) active beam-dump experiments (aBDE).  

\subsection{Tagged-photon beam fixed target experiments}
\label{S:3a}

Tagged photon-beam fixed-target experiments are hadronic-physics experiments and study mostly the excited states of mesons and baryons. These experiments can also study in-medium modifications of mesons~\cite{Trnka:2005}, and search for new forms of nuclear matter~\cite{Pheron:2012aj,PhysRevC.74.045203}, vacuum
birefringence~\cite{Hattori:2012je}, and leptophobic dark bosons~\cite{Fanelli:2016utb} and axion-like pseudo-scalar particles~\cite{Aloni:2019ruo} in Primakoff-like processes. Two types of $E\gamma E$ are suited for the Compton-like processes measurements: experiments with a solenoid (GlueX~\cite{Ghoul:2015ifw} and LEPS2~\cite{refId0}) and experiments with a forward spectrometer (FOREST~\cite{Ishikawa:2014} and LEPS~\cite{Sumihama:2006}).\\

{\bf GlueX}\\

The GlueX experiment is located at JLAB~\cite{jlab} in HallD, NewPort News, VA, USA. The photons are produced with the bremsstrahlung technique (BT) (more details about this technique can found in~\cite{Jaegle:2011sw}) and tagged between $E_\gamma =$ 9~GeV and 11.5~GeV with a resolution ($\Delta E_\gamma$) of 50~MeV. The tagged photon impinges on a 30~cm length of liquid hydrogen fixed target cell. The tagged photon flux ($\Phi_\gamma$) on target is 50~MHz. The placement of the target in chosen to get the optimal track momentum resolution ($\Delta p / p ~\sim 3\%$) and identification. Charged particles ($e^\pm$, $\pi^\pm$, $K^\pm$, and $p/\bar{p}$) and photons are tracked and/or measured between 1$^{\circ}$ and 120$^\circ$ in polar angle. Polar angles between 0$^\circ$ and 1$^\circ$ are not covered by any detectors. Charged particle(s) with a polar angle between 1$^\circ$ and 7$^\circ$  are triggering the data acquisition if 600~MeV are deposited in the Foward CALorimeter (FCAL).\\  

{\bf LEPS2}\\

LEPS2 is located at SPring8~\cite{spring8}, Sayo, Japan. The photons are produced with the Laser-Backscattering Technique (LBT)~\cite{Muramatsu:2013tla}, tagged (1.4~GeV $\leq E_\gamma \leq$~2.5~GeV and $\Delta E_\gamma$~=~12~MeV), and impinge on a 5~cm length of liquid hydrogen fixed target cell  ($\Phi_\gamma$~=~5~MHz). Charged particles and photons are tracked and/or measured between 7$^\circ$ and 120$^\circ$, and 40$^\circ$ and 120$^\circ$ in polar angle, respectively. The polar angle between 0$^\circ$ and 7$^\circ$ is covered by a $e^+e^-$ pair veto counter. Charged tracks with a momentum above 100~MeV/${\it c}$ are triggering the DAQ. The track momentum resolution on average is $\sim$ 5~$\%$. In this configuration LEPS2 has no sensitivity to particles produced in the Compton-like processes, therefore for this letter we are considering a hypothetical Multi-Wire-Drift-Chamber going down in polar angle to 1$^\circ$ as GlueX instead of 7$^\circ$ as now.\\

{\bf LEPS}\\

LEPS is also located at SPring8, Sayo, Japan, and is currently upgrading. The photons are also produced with LBT, tagged (1.4~GeV $\leq E_\gamma \leq$~2.5~GeV and $\Delta E_\gamma$~=~12~MeV), and impinge on a 5~cm length of liquid hydrogen fixed target cell  ($\Phi_\gamma$~=~5~MHz). Charged tracks with a momentum above 400~MeV/${\it c}$ and polar angle in the x-direction between -20$^\circ$ and 20$^\circ$ and y-direction between -10$^\circ$ and 10$^\circ$ are measured by the LEPS spectrometer with a track momentum resolution of 0.6$\%$. For this experiments, we will consider the rest of the 4$\pi$ solid angle to be covered by the BGOegg detector~\cite{Muramatsu:2013tla} and plastic scintillators.\\

{\bf FOREST}\\

FOREST is an experiment located at ELPH~\cite{elph}, Tohoku, Japan. The photons are produced with BT, tagged (0.8~GeV $\leq E_\gamma \leq$~1.2~GeV and $\Delta E_\gamma$~=~1~MeV), and impinge on a 5~cm length of liquid hydrogen fixed target cell ($\Phi_\gamma$~=~4.5~MHz) placed at the center of an almost 4$\pi$ setup. At a very forward angle, at a distance of 2~m from the target center, a spectrometer is positioned to measure charged tracks with momentum above 400~MeV/${\it c}$ with a polar angle in the x- and y-direction of $\pm$ 0.6$^\circ$ and $\pm$ 1.2$^\circ$, respectively. The track momentum resolution achieved is 0.6$\%$\\

Tables~\ref{tab:acc} and~\ref{tab:det} summarize the tagging system and the setup characteristics for each of the experiments we are considering.

\begin{table}
  \caption{Tagged-photon-beam characteristics}
  \centering
  \label{tab:acc}
  \begin{tabular}{llll}
   \hline
    Experiments & $\phi_\gamma$ [$\gamma/s$] & $E_\gamma$ range [GeV] & $\Delta E_\gamma$ [MeV]\\
    \hline
    GlueX &~$5 \times 10^7$ &~9~-~11.5&~50\\
    LEPS2 &~$5 \times 10^6$&~1.4~-~2.4&~12 \\                                           
    LEPS &~$5 \times 10^6$ &~1.4~-~2.4&~12 \\                                           
    FOREST &~$4.5 \times 10^6$ &~0.8~-~1.2 &~1 \\
    \hline
  \end{tabular}
\end{table}

These experiments can search for an invisible (because either the dark photon/axion-like/dark scalar is long-lived or decaying into light dark matter) or a visible (i.e. either decaying into $e^+e^-$ pair for $A'/a/\phi$ and/or $\gamma\gamma$ pair for $a/\phi$) particle. In the former, only the recoiling atomic electron and electron's bremsstrahlung photons produced in the target, air, and detector materials can be detected. In the latter, the recoiling atomic electron and/or the $e^+e^-$/$\gamma\gamma$ pair, and/or electrons' bremsstrahlung photons can be detected.

\begin{figure}
 \begin{center}
        \includegraphics[scale=0.45]{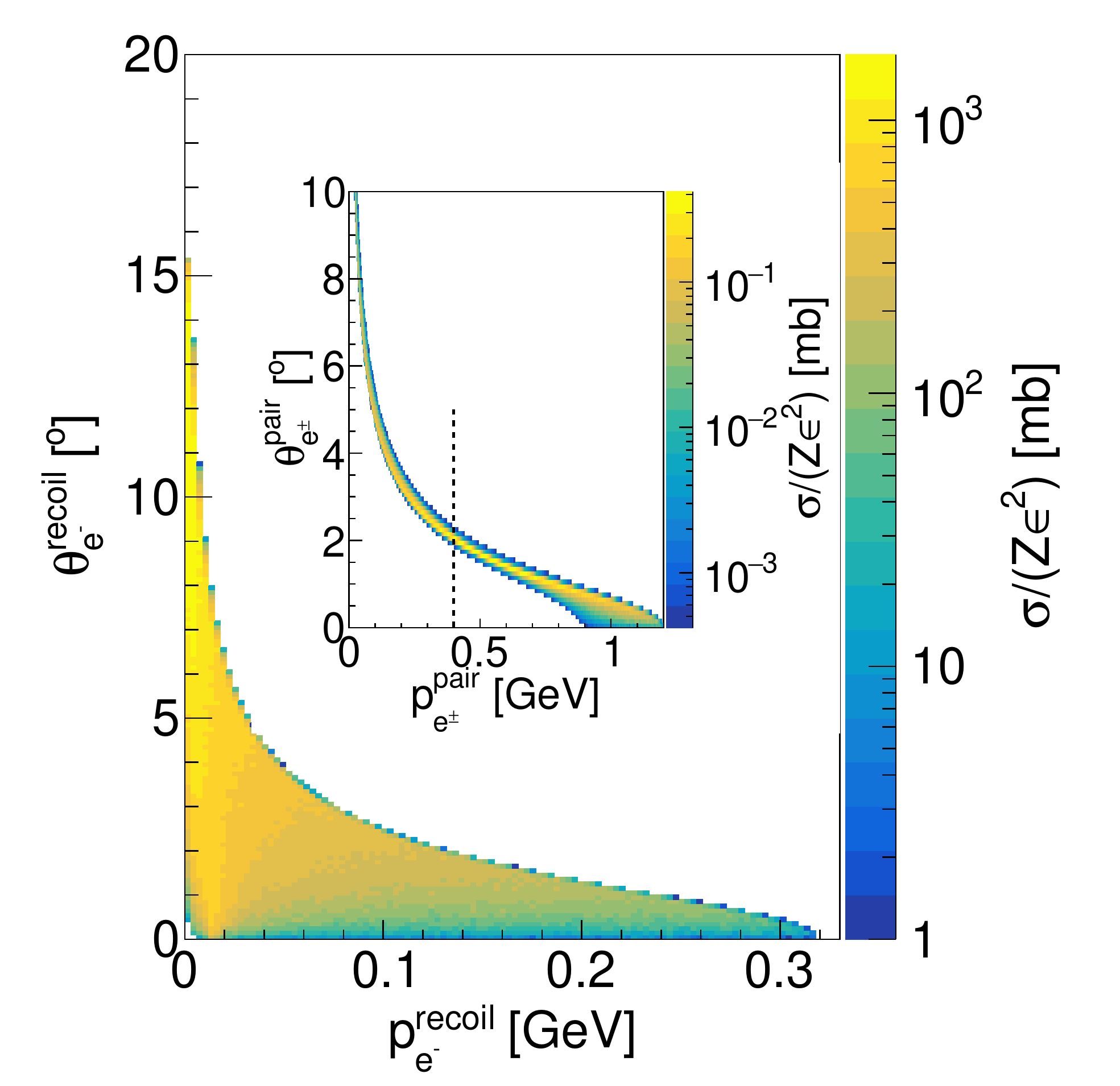}
     \caption{Recoiling atomic electron' polar angle versus momentum for a 10~MeV/${\it c}^2$ dark photon mass and FOREST experiment. The dashed black line represents the FOREST track momentum energy "threshold". The insert shows the same distribution but for the electron/positron's from the $e^+e^-$ pair.}
    \label{fig:kin}
\end{center}    
\end{figure}

The final state particles can only be detected in a very narrow polar angle range. Figure~\ref{fig:kin} shows the correlation between the recoiling atomic electron's polar angle and momentum. An energy and/or momentum threshold either due to the trigger and/or detector is/are constraining the polar angle range between 0$^\circ$ and 4.4$^\circ$, 5.7$^\circ$, 2.7$^\circ$, 5.6$^\circ$ for GlueX, LEPS2, LEPS, and FOREST, respectively. The rest of the polar range covered by the setup can be used then as a veto.  

\begin{table}
  \caption{Existing setup characteristics}
  \centering
  \label{tab:det}
  \begin{tabular}{llll}
   \hline
    Experiment~&~$\theta$~range~[$^\circ$]~&~$\Delta p / p$~[$\%$]&$p_{\mathrm{T}}^{\mathrm{track}}$ [MeV/${\it c}$]\\
    \hline
    GlueX~&~1~-~120 &~3& 50\\
    LEPS2~&~7~-~120 & 5 & 100\\
    LEPS~&~0~-~10 &~0.6& 400\\
    FOREST~&~0~-~0.6 &~0.6& 400\\
    \hline
  \end{tabular}
\end{table}

Our Monte Carlo simulation takes into account the geometrical acceptance, tagging system range and resolution, electron/positron/photon angular and momentum resolution, and electron/positron energy loss in the target and air. The following backgrounds from the following Standard Model processes are simulated: Compton ($\gamma e^- \rightarrow \gamma e^-$), pair ($\gamma N \rightarrow e^-e^+ N$ with N being the nucleon), and triplet ($\gamma e^- \rightarrow e^+e^- e^-$) photoproductions.
\begin{figure}
     \begin{center}
        \includegraphics[scale=0.45]{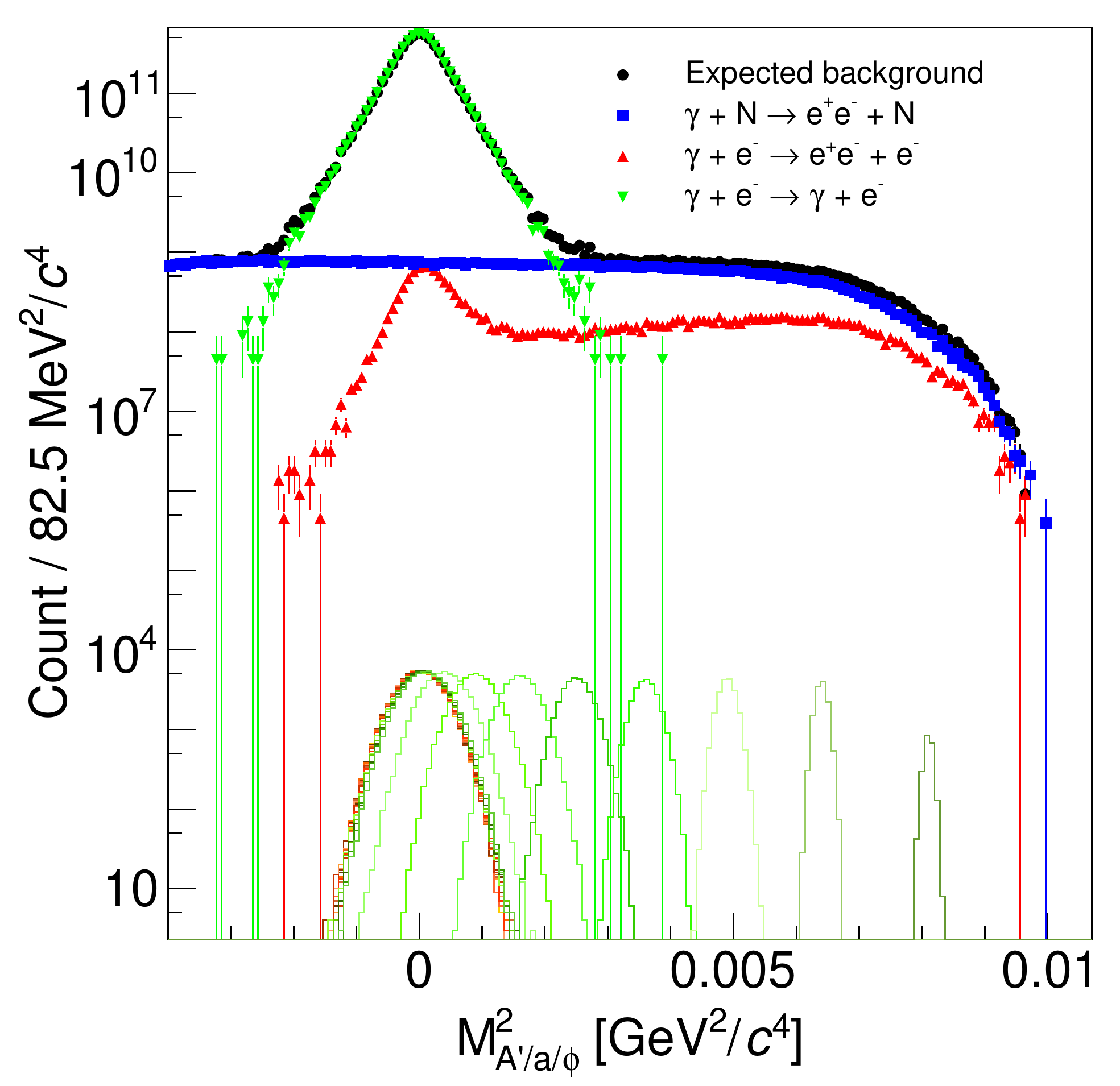}
           \caption{GlueX recoiling atomic electron missing mass squared distribution for one month of beam-time for the expected background and the expected signal, respectively. The target considered is a 30~cm long liquid hydrogen cell. The main expected contributions are the SM photoproduction of:  Compton (green point), pair (blue triangle), and triplet (red triangle). The expected signal corresponds (Gaussian-like distribution on the histogram floor) to a kinetic mixing of $10^{-4}$.}
           \label{fig:bkg_inv}
          \end{center}
\end{figure}

For the invisible case, the signature is a resonance in the distribution of the recoiling atomic electron missing mass squared, $M^2$, defined as, $M^2 = s + m_{e^-}^2 - 2 E^*_{e^-}\sqrt{s}$, where $E^*_{e^-}$ is the electron energy in the COM frame. To maximize the signal over background ratio, the following selection criteria are applied: a single track identified as an electron, the transverse momentum of the missing dark photon particle below 60, 26, 25, 10~MeV for GlueX, LEPS2, LEPS, and FOREST, respectively. The four-momentum of the missing dark photon particle ($\mbox{\bf P}_{A'}$) is defined as $\mbox{\bf P}_{A'} = {\mbox{\bf P}_{\gamma} + \mbox{\bf P}_{e^-~\mathrm{target}} -\mbox{\bf P}_{e^-~\mathrm{recoil}}}$ where ${\mbox{\bf P}_{\gamma}}$, ${\mbox{\bf P}_{e^-~\mathrm{target}}}$, and ${\mbox{\bf P}_{e^-~\mathrm{recoil}}}$
are the four-momenta of the incident photon, the initial state electron 
(assumed to be at rest), and the recoiling atomic electron, respectively. In addition, for LEPS and FOREST, the recoiling atomic electron transverse momentum is above 1.85~MeV. Finally, the polar angle of the missing dark photon particle is below 0.55$^\circ$, 2.4$^\circ$, 5$^\circ$, and 10$^\circ$ for GlueX, LEPS2, LEPS, and FOREST, respectively.
Figure~\ref{fig:bkg_inv} shows the remaining expected GlueX recoiling atomic electron missing mass squared distribution after all selection criteria described above are applied for the background and signal for different dark photon mass hypothesis and a kinetic mixing of $10^{-4}$. For the solenoid experiments, GlueX and LEPS2, the addition of a ``gamma" and ``$e^+e^-$ pair" veto detector covering the polar angle between 0$^\circ$ and 1$^\circ$ could reduce the background up to none if the veto detector is 100$\%$ efficient as the backgrounds considered will always produce an electron or a photon in this polar range. For the spectrometer experiments, LEPS and FOREST, the background can be strongly reduced if the inside exterior wall of the spectrometer is instrumented to measure the electrons/positrons with a momentum below 400~MeV/${\it c}$.

For the visible case, as function of the mass of the particle candidate, we get a better result by either measuring the recoiling atomic electron, the $e^+e^-$ pair, or both of them simultaneously. GlueX and LEPS2 can either measure the recoiling atomic electron or the $e^+e^-$ pair, while LEPS and FOREST can measure both simultaneously. The selection criteria are the same as for the invisible case when measuring the recoiling atomic electron, except that up to two additional tracks are allowed, identified as an electron or positron and the sum of all tracks charge being 0 or -1 or -2. When the $e^+e^-$ pair can be reconstructed, the following selection criteria are used: transverse momentum of the $e^+e^-$ pair and recoiling $e^+e^-$ pair missing particle between 15~MeV/$\it c$ and 50~MeV/$\it c$, 4~MeV/$\it c$ and 50~MeV/$\it c$, 2~MeV/$\it c$ and 21~MeV/$\it c$ for GlueX, LEPS2/LEPS, and FOREST, respectively. For GlueX and LEPS2, in addition, the recoiling $e^+e^-$ pair missing particle energy is above 600~MeV and 200~MeV, respectively. If the recoiling atomic electron and the $e^+e^-$ pair are reconstructed simultaneously, the events with an energy and mass difference between the initial states and final states of 30~MeV/${\it c}^2$ and 2~MeV/${\it c}^2$ are selected, respectively. 

In principle, GlueX can measure the $\gamma\gamma$ pair invariant mass but the FCAL energy and angle resolutions are not competitive compared to the measurement of the recoiling atomic electron.   

We are only considering the background produced in the target by the photon beam. The background produced along the beam-line can be estimated by an empty target run. 

\begin{figure}
 \begin{center}
        \includegraphics[scale=0.45]{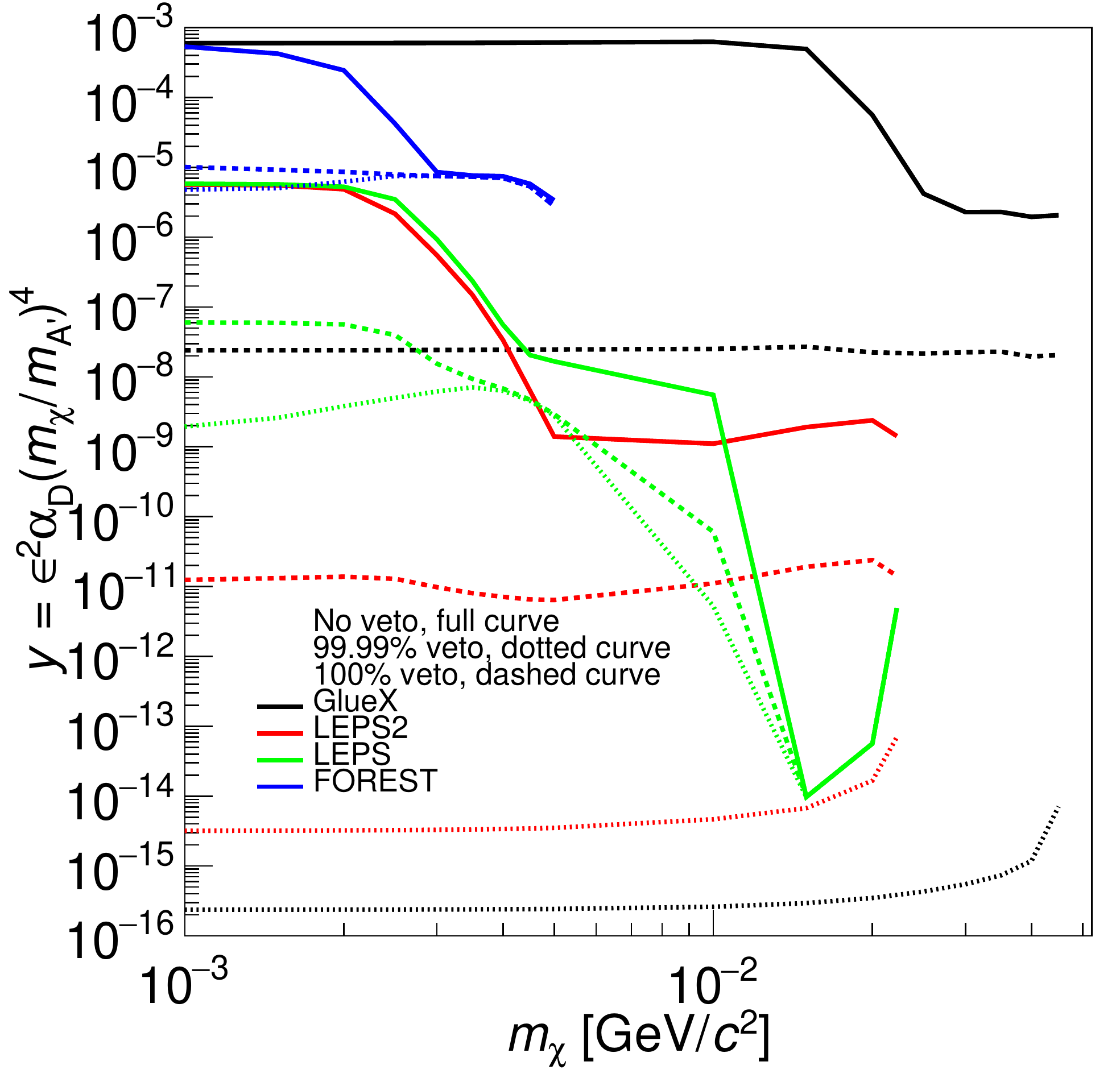}
          \caption{90$\%$ CL expected sensitivity on the dimensionless interaction strength $y = \epsilon^2\alpha_D(m_\chi/m_{A'})^4$ (for the hypothesis of $\alpha_D=0.5$ and $m_{A'}=3m_\chi$) versus the light dark matter mass for GlueX (black curve), LEPS2 (red curve), LEPS (green curve), and FOREST (blue curve) experiments. Dashed curve: ``as is" i.e. without modifying the detector except for LEPS2 where the MWDC minimum polar coverage has been changed from 7$^o$ to 1$^o$. Full and dotted curves: addition of an hypothetical veto detector which is 99.99$\%$ (dashed curve) and 100$\%$ (dotted curve) efficient covering the polar angle between 0$^o$ and 1$^o$ for GlueX and LEPS2, and inside the spectrometer for LEPS and FOREST.} 
          \label{fig:comp_inv}
          \end{center}
\end{figure}

\begin{figure}
 \begin{center}
        \includegraphics[scale=0.45]{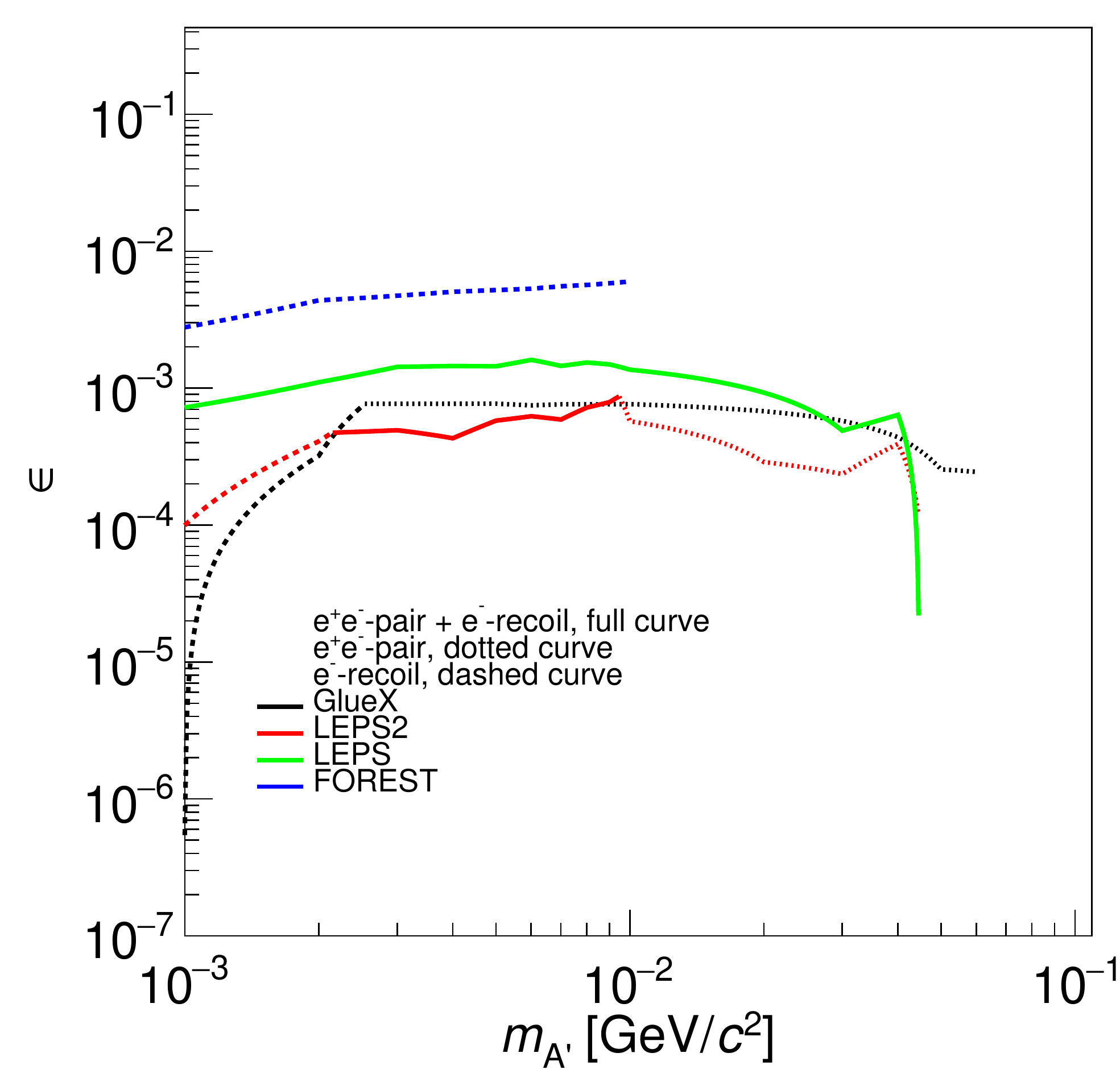}
          \caption{90$\%$ CL expected sensitivity on $\epsilon$ versus the dark photon mass for GlueX (black curve), LEPS2 (red curve), LEPS (green curve), and FOREST (blue curve) experiments. Dotted curve: recoiling missing mass analysis with one to three tracks measured. Full and dashed curves: invariant mass analysis from two and three tracks measured, respectively.}
          \label{fig:comp_vis}
          \end{center}
\end{figure}

The sensitivity is calculated for different background scenarios: 100$\%$ and 99.99$\%$ efficient veto detector, and no veto detector. An unbinned extended log-likelihood fit in one-dimension, is done by using a Gaussian PDF with a fixed width for the signal and a Crystal Ball function~\cite{Oreglia:1980cs} plus a 5th order Polynomial for the background probability density function (PDF) to estimate the signal yield. The signal width of the missing mass squared of the dark photon varies between 360 and 68~MeV$^2$/${\it c}^4$, 9 and 5~MeV$^2$/${\it c}^4$, 7 and 4~MeV$^2$/${\it c}^4$, and 5 and 4~MeV$^2$/${\it c}^4$ for the GlueX, the LEPS2, the LEPS, and the FOREST experiments, respectively. The signal width of the $e^+e^-$ pair invariant mass varies between 70~keV/${\it c}^2$ and 20~MeV/${\it c}^2$, 70 and 350~keV/${\it c}^2$, 60 and 200~keV/${\it c}^2$, and 60 and 70~keV/${\it c}^2$ for the GlueX, the LEPS2, the LEPS, and the FOREST experiments, respectively. The fit is repeated for a dozen different possible masses of the signal. We compute an upper limit (UL) at 90$\%$ confidence level (C.L.) on the signal yield by integrating the likelihood function.  The systematic uncertainty is accounted for in the limit calculation  by convolving the  likelihood  with a Gaussian function, which has a width equal to the total systematic uncertainty. For all the experiments, we are considering the systematic uncertainty of the photon flux to be 10$\%$ and the detection efficiency to be 40$\%$. The upper limit on the number of observed events is also equaled to:
\begin{equation}
   N_{\mathrm{obs}}^{\mathrm{90\%~CL}} = \sigma \cdot {\cal L} \cdot \varepsilon \cdot \mathrm{BR} 
\end{equation}
where $\sigma$, $\varepsilon$, and $\mathrm{BR}$ correspond to the theoretical cross section (Equation~\ref{eqn:11}), the detection efficiency, and the branching ratio, respectively. The luminosity due to photon with an energy above the production threshold is expressed as:  
\begin{equation}
{\cal L} =  \frac{{\cal N_A}}{A} \cdot \rho \cdot l_{\mathrm{target}} \cdot \Phi_\gamma \cdot \Delta t \label{eqn:lum}
\end{equation}
where $\cal N_A$, $A$, $\rho$, $l_{\mathrm{target}}$, $\Phi_\gamma$, and $\Delta t$ are the Avogadro number, the molar mass, the target density, the target length, the photon flux, and the beam time, respectively. The recoiling atomic electron detection efficiency varies between 13 and 0.8$\%$, 28 and 7$\%$, 20 and 0.04$\%$, and 1 and 2$\%$ for the GlueX, the LEPS2, the LEPS, and the FOREST experiments, respectively. The $e^+e^-$ pair detection efficiency varies between 1 and 0.01$\%$, 3 and 0.01$\%$, 20 and 0.5$\%$, and 10 and 0.2$\%$ for the GlueX, the LEPS2, the LEPS, and the FOREST experiments, respectively.

Figure~\ref{fig:comp_inv} and ~\ref{fig:comp_vis} show the expected 90$\%$ C.L. sensitivity to the dimensionless interaction strength $y$ versus the light dark matter mass and to the kinetic mixing between the dark photon and the SM photon versus the dark photon mass for GlueX, LEPS2, LEPS, and FOREST.  

\subsection{Beam-dump experiments}
\label{S:3b}

In the beam-dump experiments~\cite{Riordan:1987aw,Bjorken:1988as,Bross:1989mp,Davier:1989wz,Konaka:1986cb,Andreas:2012mt,Battaglieri:2017qen,Marsicano:2018glj,Marsicano:2018jd}, an incident electron of fixed energy of several 100's of MeV to several 100's of GeV produces an electromagnetic shower in a heavy $Z$ target (dump) of several radiation lengths, $X_0$. Typically after the dump, there is a (concrete and/or bedstone) radiation shield of length, $L_{\mathrm{sh}}$, followed by an open space or tunnel of diameter of few cm's to few 10's of cm and of length, $L_{\mathrm{dec}}$, the so-called decay length. The detectors are placed at the end of the decay length. These experiments are typically looking for long-lived particles ($A'/a/\phi$) decaying into SM particles in the $L_{\mathrm{dec}}$ region and/or invisible particles (dark matter, $\chi$) produced directly in the dump. 

We simulate with the Geant4 simulation package~\cite{Agostinelli:2002hh,Allison:2006ve,Asai:2015xno} the electromagnetic shower occurring in the dumps for the experiments: E141~\cite{Riordan:1987aw}, E137~\cite{Bjorken:1988as}, E774~\cite{Bross:1989mp}, Orsay~\cite{Davier:1989wz}, KEK~\cite{Konaka:1986cb}, and the planned beam-dump experiment BDX~\cite{Battaglieri:2017qen,Marsicano:2018glj,Marsicano:2018jd}. For all experiments, the geometry and the materials of the dump and the decay length region are implemented in Geant4. For all photons with an energy above 20~MeV, we store the photon energy and direction, and the Geant4 production and decay vertices (the difference between these two vertices corresponds to the photon Geant4 interaction length, $l_\gamma$). The luminosity can be written as:
\begin{equation}
  {\cal L} =  \frac{{\cal N_A}}{\sum_{i} w_i A_i} \cdot \sum_{i} w_i \rho_i \cdot l_\gamma \cdot N_\gamma  
\end{equation}
where $w_i$ is the proportion of the different atomic element composing the dump. $N_\gamma$ is the number of photons produced in the dump with an energy above the production threshold. This number is estimated with Geant4 as follows: $N_\gamma = \mathrm{EOT} \cdot N_\gamma^{\mathrm{Geant4}}~/~\mathrm{EOT}^{\mathrm{Geant4}}$ with $\mathrm{EOT}$ the number of electron on target (Table~I of~\cite{Andreas:2012mt} of~\cite{Marsicano:2018glj}). $\mathrm{EOT^{Geant4}}$ and $N_\gamma^{\mathrm{Geant4}}$ are the simulated number of electrons on target and the number of photons produced in the dump by Geant4, respectively. In a dump composed of a single element, the photon interaction length is equal to $\frac{9}{7}X_0$, where the ratio $\frac{9}{7}$ arises from the fact that the photon does not interact as often as the electron/positron, and originates from the cross-section ratio between the SM bremsstrahlung and the SM pair productions. This is a valid approximation for $E_\gamma \geq$~20~MeV.

\begin{table}
  \caption{BDE detector main characteristics}
  \centering
  \label{tab:BDEdet}
  \begin{tabular}{llll}
   \hline
    Exp.~&~ Section ~&~$E_{\mathrm{threshold}}^{\mathrm{det}}$~&~$E_{\mathrm{range}}^{\mathrm{det}}$~\\
    &~[cm$^2$]~&~[GeV]~&~[GeV]~\\
    \hline
    E774~&~10~$\times$~10~&~2.75~&~272.5~\\
    E141~&~7.5~$\times$~7.5~&~6.3~&~1.8~\\
    Orsay~&~15~$\times$~15~&~1.1~&~0.5~\\
    KEK~&~6.4~$\times$~11.2~&~0.1~&~2.4~\\
    E137-I~&~200~$\times$~300~&~1~&~19~\\
    E137-II~&~300~$\times$~300~&~1~&~19~\\
    BDX~&~50~$\times$~40~&~0.5~&~10.5~\\
    \hline
  \end{tabular}
\end{table}

Then, for each Geant4 photons with an energy above the production threshold and 10000 COM polar angle between 0$^\circ$ and 180$^\circ$, we calculate the Compton differential cross section for a given mass and $\epsilon$, and the process kinematics for 10000 dark photons by assuming a phase space COM azimuthal angle. Each dark photon is boosted into the laboratory frame and decays into two electrons which follow in the dark photon rest frame a $\frac{3}{2} + \mathrm{cos}(\theta)$ distribution corresponding to a mother with spin 1. The dimensions of the dump, shield, and decay regions can be found as well in Tables~I of~\cite{Andreas:2012mt} and of~\cite{Marsicano:2018glj}. If the dark photon decays in the $L_{\mathrm{dec}}$ region and the electron and/or positron have a trajectory pointing within the geometrical detector acceptance, they are then sent back to Geant4 to simulate the energy loss in air in the $L_{\mathrm{dec}}$ region. If the electron and/or positron, at the detector entrance cross-section, have an energy above the energy threshold and within the measurable energy range, the event is accepted. In Table~\ref{tab:BDEdet}, we list the detector vertical cross-section, energy threshold ($E_{\mathrm{threshold}}^{\mathrm{det}}$), and energy range ($E_{\mathrm{range}}^{\mathrm{det}}$), we use in our Monte Carlo simulation. The detector response is not simulated. 

The number of expected event is expressed as follows:
\begin{equation}
     N_{\mathrm{expected}} =  \sigma \cdot {\cal L} \cdot \mathrm{BR} \cdot \varepsilon \cdot \sum_{l = L_{sh}}^{L_{sh} + L_{dec}} P(l)\Delta l
\end{equation}
where $P(l)$ and $\Delta l$ are the decay length probability and the decay length step, respectively. $P(l)$ is defined as followed: $P(l) = \frac{1}{l_{A'}}e^{-\frac{l}{l_{A'}}}$ with $l$ and $l_{A'}$ the decay length and the mean decay length, respectively. $l_{A'}$ is equal to $l_{A'} = \gamma \beta \tau$ with $\gamma$ the Lorentz boost and $\beta = \sqrt(1 - \frac{1}{\gamma})$. $\tau$ is the decay time and is equal to $\tau = 1 / \Gamma$. $\Gamma$ is the decay width and is taken from~\cite{Bjorken:2009mm,Essig:2009nc,Essig:2010xa}:
\begin{equation}
\Gamma =  \frac{\alpha \epsilon^2}{4} \left( 1 + 2 \frac{m_l^2}{m_{A'}^2} \right) \sqrt{1 - 4 \frac{m_l^2}{m_{A'}^2}}   
\end{equation}

The contour plot is done for $N_{\mathrm{expected}} / N_{\mathrm{observed}}^{\mathrm{90\%~CL}}~\geq~1$
if there are zero expected background and zero observed event $N_{\mathrm{observed}}^{\mathrm{90\%~CL}}~=~2.3$. The experiments Orsay, KEK, and E137 are expected to measure zero background and observe zero event while the experiments E774 and E141, are expected to measure non-zero background and observe 1126 and 0 events, respectively. We calculate the 90$\%$ upper limit on the number of observed event with a Bayesian inference method with the use of Markov Chain Monte Carlo~\cite{Cadwell:2009} and find for the experiments E774 and E141: 2621.23 and 13.8, respectively.

\begin{figure}
 \begin{center}
        \includegraphics[scale=0.45]{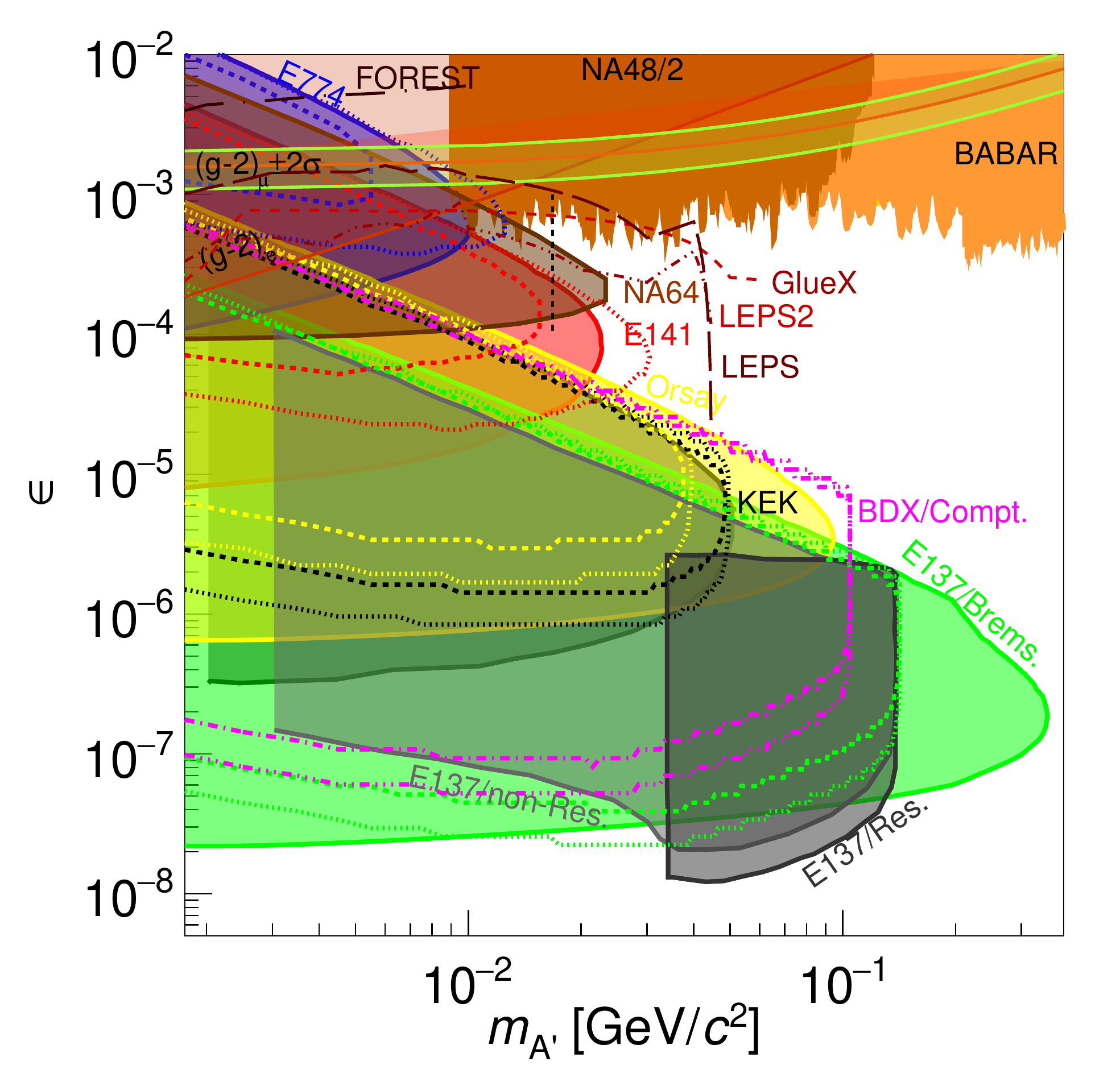}
 
          \caption{90$\%$ C.L. contour limit on $\epsilon$ versus the visible dark photon mass for the Compton-like process compared to Bremsstrahlung- and meson-like processes. The dotted curves correspond to the Compton process contour limits of the past beam-dump experiments: E774, E141, KEK, Orsay, E137. The dashed dotted curve corresponds to the expected Compton process sensitivity of the future beam-dump experiment: BDX. The medium and long dashed curves correspond to the expected Compton process sensitivity of existing real photon beam experiments: LEPS2/E949 and GlueX; with and without the expected backgrounds included. All the filled areas correspond to Brems-strahlung- (BDE~\cite{Riordan:1987aw,Bjorken:1988as,Bross:1989mp,Davier:1989wz,Konaka:1986cb,Andreas:2012mt}, BABAR~\cite{Lees:2014xha}, and NA64~\cite{Banerjee:2018vgk}) and meson-like processes (NA48/2~\cite{Batley:2015lha}).}
          \label{fig:bd_epsilon}
          \end{center}
\end{figure}

\begin{figure}
 \begin{center}
    \includegraphics[scale=0.45]{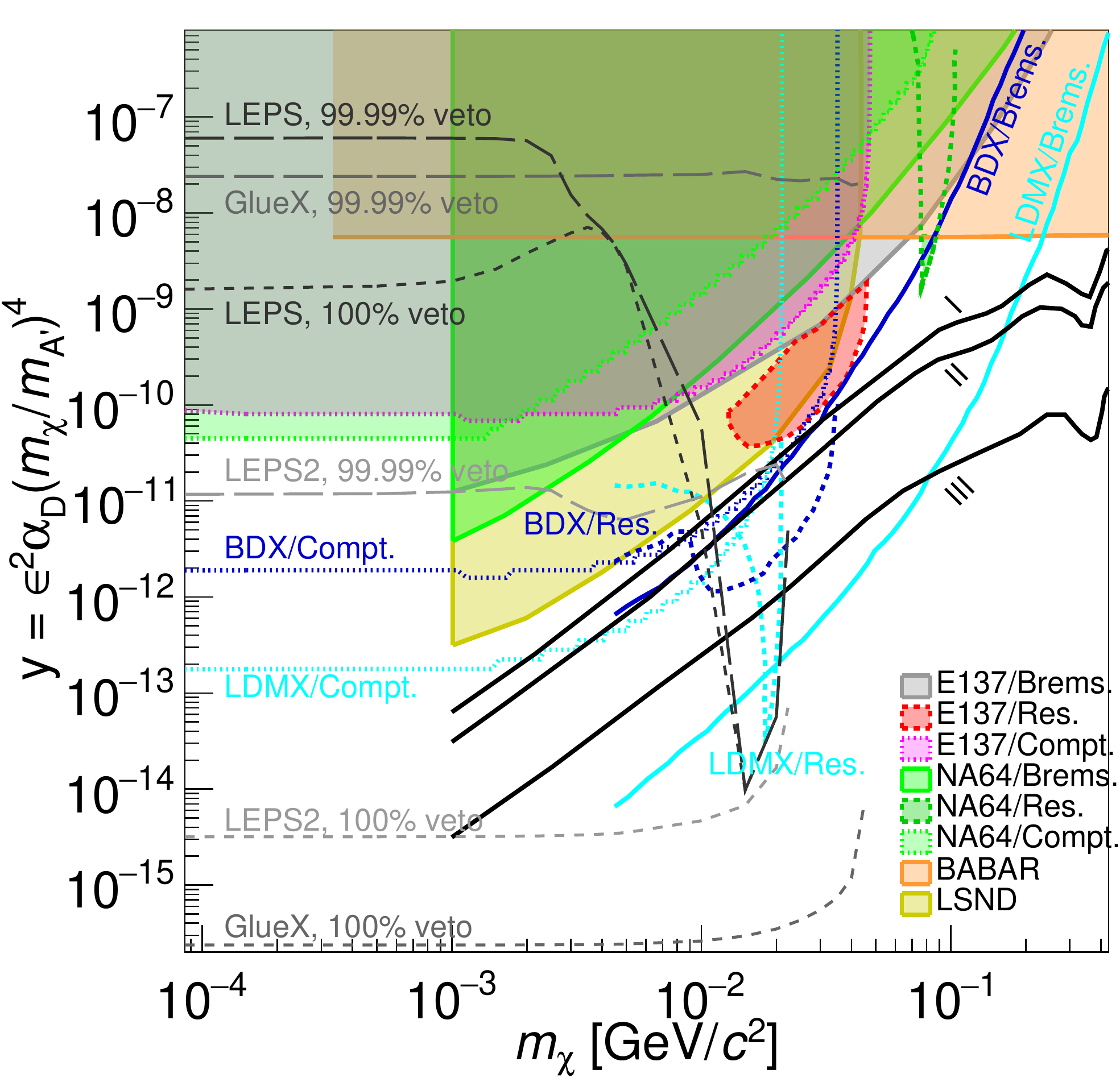}
          \caption{90$\%$ C.L. contour limit on the dimensionless interaction strength $y = \epsilon^2\alpha_D(m_\chi/m_{A'})^4$ (for the hypothesis of $\alpha_D=0.5$ and $m_{A'}=3m_\chi$) versus the on-shell light dark matter mass for the Compton-like process compared to the Bremsstrahlung-like process. All filled areas correspond to bremsstrahlung process (E137~\cite{Bjorken:1988as,Marsicano:2018glj}, BABAR~\cite{Lees:2017lec}, NA64~\cite{Banerjee:2017hhz}, and LNSD~\cite{Auerbach:2001wg}) except for the pink and light  green areas which correspond to the Compton E137 and NA64 contour limits, respectively. The three solid black curves represent the thermal relic target for different cases on the nature of the light dark matter: elastic and inelastic scalar (I), Majorana fermion (II), and pseudo-Dirac fermion (III).} 
          \label{fig:bd_ldm}
          \end{center}
\end{figure}

The experiment E137 and the future BDX experiments can also detect dark matter through the dark matter scattering off electron process: $\chi e^- \rightarrow \chi e^-$~\cite{Izaguirre:2013uxa,Marsicano:2018glj,Marsicano:2018jd}. The light dark matter differential cross-section scattering off a free electron, $d\sigma_{\chi e}/dT_R$ can be expressed as~\cite{Izaguirre:2013uxa}:
\begin{equation}
\frac{d\sigma_{\chi e}}{dT_R} = 4\pi \alpha \alpha_D \epsilon^2   m_e \frac{4 m_e m^2_\chi T_R + \left[m^2_\chi +m_e(T-T_R)\right]^2}{(m^2_{A'}+2m_e T_R)^2(m^2_\chi+2m_e T)^2} \;  ,
\label{eqn:diff_ldm}
\end{equation}
where $T$ and $T_R$ are the light dark matter and the recoiling $e^-$ kinetic energies, respectively. The dark matter particle is produced in the dump via the Compton-like process: $\gamma e^- \rightarrow A' e^-$ and $A' \rightarrow \chi\bar{\chi}$. The electron off which the dark matter particle is scattering off is an atomic electron of the detector material (which is located after the decay region). To estimate the number of expected events, we followed the same methodology and used the same parameters as in~\cite{Marsicano:2018glj,Marsicano:2018j}. The only difference is that we used the dark photons produced by the C-reactions instead of the B-reactions.
The number of light dark matter detected, $N^{s}_{\chi-e}$, can be written as(~\cite{Marsicano:2018glj}-equation 3):
\begin{equation}
N^{s}_{\chi-e}= N_{\chi\overline{\chi}} \, n_e L_{det} \, \sigma^*_{\chi e} \varepsilon_{s} \; ,
\end{equation}
where $N_{\chi\overline{\chi}}$(~=~2$N_{A'}$), $n_e$, $L_{det}$, $\sigma^*_{\chi e}$, $\varepsilon_{s}$ are the number of light dark matter produced on-shell via the C reaction, the detector electron density taken from Table~I of~\cite{Marsicano:2018glj}, the detector length taken from Table~I of~\cite{Marsicano:2018glj}, the integral of the light dark matter scattering off a free electron differential cross section (Eq.~\ref{eqn:diff_ldm}) from the detector energy threshold to maximum measurable energy, and the detector detection efficiency, respectively. 

\subsection{Active beam-dump experiments}
\label{S:3c}

The active beam-dump experiments are either measuring the energy deposited in the target (NA64~\cite{Banerjee:2016tad,Banerjee:2017hhz,Banerjee:2018vgk}) or measuring the recoil momentum (LDMX~\cite{Akesson:2018vlm}). Typically, when the energy deposited is measured in the target, the target length is of several radiation length. While when the momentum is measured, the target length is of several 100's $\mu m$. We estimate again the number of expected events by following the  methodology and setup parameters described in~\cite{Marsicano:2018glj,Marsicano:2018j}. In particular, we are using equations 5 of~\cite{Marsicano:2018glj} which is expressing the number of dark photon detected, $N^{s}_{A'}$: 
\begin{equation}
N^{s}_{A'} = \varepsilon_{s} \int^{T_0}_{E^{CUT}_{miss}} N_{A'} (E) dE \; .
\end{equation}
where $\varepsilon_{s}$, $T_0$, $E^{CUT}_{miss}$ are the detector detection efficiency, the incident electron kinetic energy, the detector energy threshold. The main difference between the BDE and the aBDE is that for the latter the number of event detected is not suppressed by $\alpha_D$ as in the BDE.

\section{Results and discussions}

Figures~\ref{fig:bd_epsilon} and~\ref{fig:bd_ldm} summarizes the current best results and projected sensitivities for the B-reactions as well as the new results and expected sensitivities for the C-reactions on the kinetic mixing between the dark photon and the SM photon, and on the dimensionless interaction strength $y$, respectively. For all BDE and aBDE, the limits and expected sensitivities obtained with the B-reactions supplant the ones obtained with the C-reactions. For a dark photon mass above 10~MeV/${\it c}^2$, the B-cross-section is roughly an order of magnitude larger than the C-cross-section, therefore, if one combines the two limits due to the two processes, we are expecting the regions of sensitivities to be extended to higher masses for E774 and E141 and lower kinetic mixing values for the other BDE as illustrated by the dashed curves in Figure~\ref{fig:bd_epsilon}. For the aBDE, presumably if the two recoil electrons coming from the SM bremsstrahlung process and the dark Compton process can be measured simultaneously i.e. if the photon can be tagged, the energy threshold could be decreased, and therefore the expected sensitivities to the C-reaction could be increased. The expected sensitivities of the tagged photon fixed target experiments show that they could potentially scan the Be anomaly for the visible case and are competitive for the measurement of the invisible case provided that they are tuned for these measurements.

Another advantage of the tagged photon beam fixed target experiments is that the circularly polarized photon beam and the transversely polarized target may open the possibility to measure the spin of the dark particle with the so-called double polarization $E$ observable and therefore distinguish a scalar- or a pseudo-scalar particle from a vector particle and vice versa depending the orientation of the target polarization.

\section{Conclusion}
\label{S:4}

We highlight the experimental possibilities offered by the Compton-like processes for the search for the dark photon, the axion-like pseudo-scalar, the dark scalar mediator, and the light dark matter. We show that the beam-dump experiments and possibly the active beam-dump experiments should have increased sensitivities at certain region of parameter-space. We also show that the tagged photon-beam fixed-target experiments are competitive compared to existing and planned electron beam fixed target experiments and could potentially extract a limit on the dimensionless interaction $y$ for dark matter with a mass lighter than 35 MeV/${\it c}^2$ up to 10$^{-16}$ with one month beam time and a hydrogen target.

\section{Acknowledgement}

We would like to thank Jason Kumar for pointing to problem 5.5 in the Peskin and Schroder's book ``An Introduction To Quantum Field Theory", John Yelton and Pierre Sikivie for kindly reviewing this article, Yimming Zhong for sharing and explaining how to use the Madgraph models for the bremsstrahlung-like processes, the GlueX, the LEPS2, the LEPS, the FOREST, the BDX, the NA64, and the LDMX Collaborations (in particular from these Collaborations Andrea Celantano, Volker Crede, Sean Dobbs, Bertrand Echenard, Sergei Gninenko, Takashi Nakano, Masayuki Niiyama, Norihito Muramatsu, Justin Steven, and Atsushi Tokiyasu) for all their comments and discussions. The work of Igal Jaegle was supported by the Office of High Energy Physics of the U.S. Department of Energy under grant contract number DE-SC0009824. The work of Sankha S. Chakrabarty was supported in part by the U.S. Department of Energy under grant DE-SC0010296 and by the Dissertation Fellowship from the University of Florida.

\bibliography{DarkCompton.bib}

\section{Appendix}
\label{sec:appendix}
The differential cross-section of the Compton-like process $e^- (p_1) \; \gamma(p_2) \rightarrow e^- (k_1) \; A'(k_2)$ in the center of momentum (COM) frame is given by:
\begin{equation}
\left( \frac{d\sigma}{d\Omega} \right)_{\text{COM}} = \frac{\alpha^2 \epsilon^2}{8s} \frac{1}{(1-x_m^2)} \frac{k}{\sqrt{s}} (T_1 + T_2 + T_3) \label{diff_xs}
\end{equation}
where $\alpha = \frac{e^2}{4\pi}$ is the fine structure constant, $\sqrt{s} = E_{\text{com}}$ is the energy in center of momentum frame, $m_e =x_m\sqrt{s}$ is the electron mass and $M = x_M\sqrt{s}$ is the dark photon mass and
\begin{widetext}
\begin{eqnarray}
\frac{k}{\sqrt{s}} &&= \frac{1}{2} \sqrt{(1+x_m^2 - x_M^2)^2 - 4x_m^2} \nonumber \\
\left( \frac{u-m_e^2}{s} \right) &&= x_M^2 - \frac{1}{2} (1+x_m^2)(1-x_m^2 + x_M^2) - \frac{k}{\sqrt{s}} (1-x_m^2) \cos \theta \nonumber \\ 
T_1 &&= \frac{16}{(1-x_m^2)^2} \left[ 2x_m^4 + x_m^2 (1-x_m^2) + x_m^2 x_M^2 - \frac{1}{2}(1-x_m^2) \left( \frac{u-m_e^2}{s} \right) \right] \nonumber \\
T_2 &&= \frac{16}{(1-x_m^2)\left( \frac{u-m_e^2}{s} \right)} \left[ x_m^2(1-x_m^2) + x_m^2 \left( \frac{u-m_e^2}{s} \right) + 4x_m^4 + x_M^2 (1 - x_m^2 - x_M^2 + \left( \frac{u-m_e^2}{s} \right))\right] \nonumber \\
T_3 &&= \frac{16}{\left( \frac{u-m_e^2}{s} \right)^2} \left[ 2x_m^4 + x_m^2\left( \frac{u-m_e^2}{s} \right) + x_m^2 x_M^2 - \frac{1}{2}(1-x_m^2)\left( \frac{u-m_e^2}{s} \right) \right] .
\end{eqnarray}
\end{widetext}

\end{document}